\documentclass [prl,superscriptaddress, showpacs, twocolumn]{revtex4-1}
\usepackage{graphicx}
\usepackage{dcolumn}

\makeatletter

\newcommand{\Rmnum}[1]{\expandafter\@slowromancap\romannumeral #1@}
\makeatother

\begin{document}

\title{A possible chiral spin-liquid phase in non-centrosymmetric $R$BaCo$_4$O$_7$}

\author{D.D. Khalyavin}
\affiliation{ISIS facility, STFC Rutherford Appleton Laboratory, Chilton, Didcot, Oxfordshire, OX11-0QX,United Kingdom}
\author{P. Manuel}
\affiliation{ISIS facility, STFC Rutherford Appleton Laboratory, Chilton, Didcot, Oxfordshire, OX11-0QX,United Kingdom}
\author{L.C. Chapon}
\affiliation{Institut Laue-Langevin, BP 156X, 38042 Grenoble, France}

\date{\today}

\begin{abstract}
Based on a symmetry approach, we propose a possible explanation of the weak ferromagnetic component recently observed in YBaCo$_3$FeO$_7$ (Valldor et al. Phys Rev B, $\bf {84}$ 224426 (2011)) and other isostructural compounds in the high-temperature spin-liquid phase. Due to the polar nature of their crystal structure, a coupling between time-odd scalar spin chirality which we suggest as the primary order parameter and macroscopic magnetization is possible as follows from the general form of the appropriate free-energy invariant. The deduced pseudoproper coupling between both physical quantities provides a unique possibility to study the critical behaviour of the chiral order parameter.
\end{abstract}


\maketitle

\begin{figure}[t]
\includegraphics[scale=1.0]{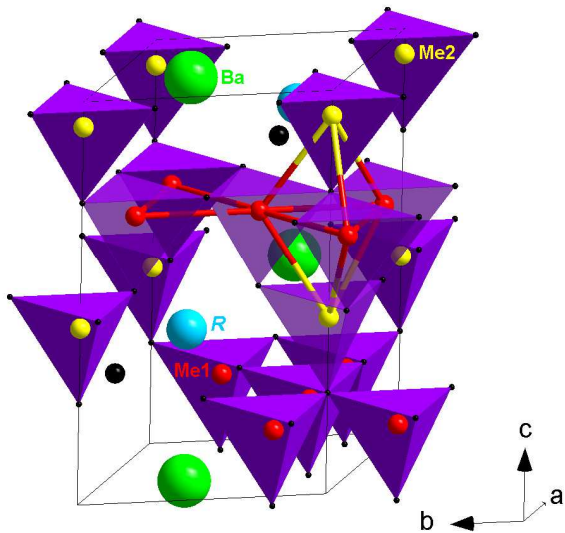}
\caption{(Color online) Schematic representation of the hexagonal $P6_3mc$ crystal structure of $R$BaMe$_4$O$_7$ as a corner-shared MeO$_4$ tetrahedral network. Metals ions in Kagome and triangular layers are denoted as Me1 (red) and Me2 (yellow), respectively. Solid lines represent exchange bonds.}
\label{fig:F1}
\end{figure}

\begin{figure}[t]
\includegraphics[scale=0.91]{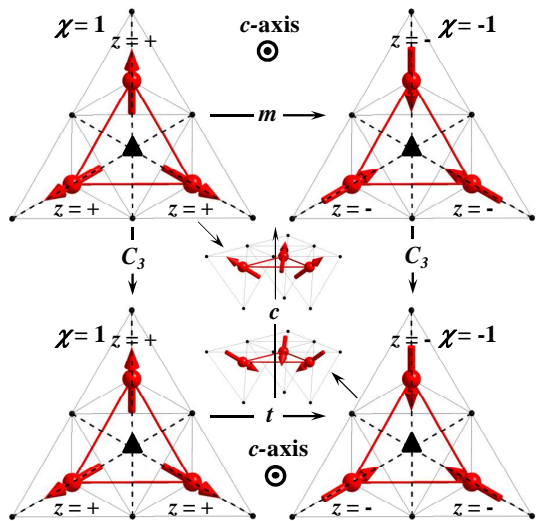}
\caption{(Color online) Transformational properties of three non-coplanar spins in a triangle (for simplicity, we arbitrary took these spins to create equal angles between each other but actually the angles can be different). The scalar chirality, $\chi $, changes sign under the mirror reflection, $m$, and time-inversion, $t$, but does not change it under the rotation $C_3$. Intersections of the mirror planes of the triangle with the plane containing it are shown as dot-lines. The signs of $z-$component for each spin are shown as "+ "or "-"}
\label{fig:F2}
\end{figure}

\begin{figure}[t]
\includegraphics[scale=1.43]{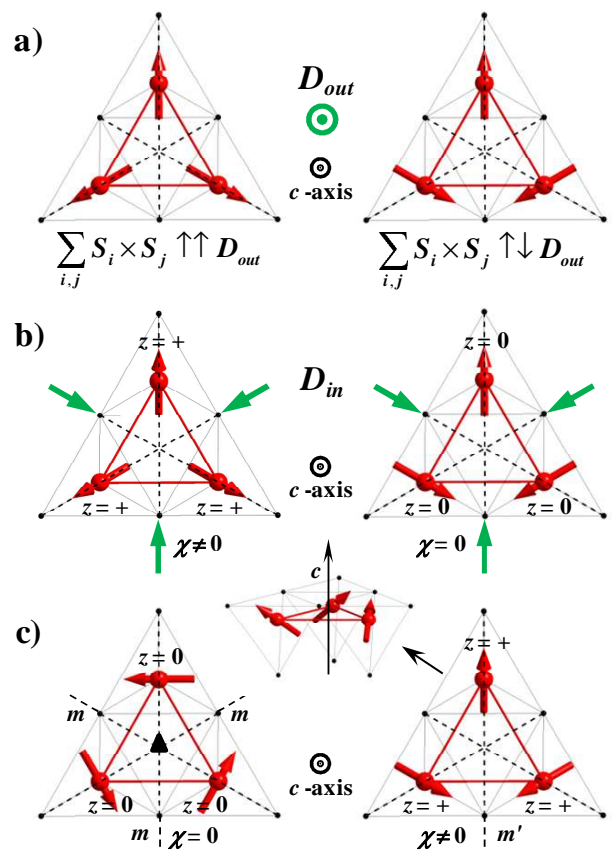}
\caption{(Color online) Two variants of coplanar spin configurations in a triangle with parallel $(\uparrow \uparrow )$ and antiparallel $(\uparrow \downarrow )$ directions of the vector chirality and $\bf {D_{out}}$ component. The latter was arbitrary chosen to be in the up-direction (a). Impact of the in-plane $\bf {D_{in}}$ component on these spin configurations (b). Spin configuration forbidding (left) and allowing (right) $z$-component and scalar chirality to be non-zero (c).}
\label{fig:F3} 
\end{figure}

\indent The concept of symmetry and its spontaneous breaking at a phase transition is a central topic of condensed matter physics. Invariance of physical properties of crystals under spatial operations (such as rotations and reflections) and time-reversal is the basis for their comprehensive description and classification. Time-reversal symmetry is the crucial ingredient to understand properties and describe phase transitions in magnetic systems with long-range spin-dipole ordering. Combination of this symmetry with the spatial operations form a powerful tool of magnetic symmetry based on which such fundamental phenomena as antisymmetric exchange \cite{ref:1} piezomagnetism \cite{ref:2} and linear magnetoelectric effect \cite{ref:3} have been predicted.\\
\indent Besides conventional magnets exhibiting spontaneous breaking of time-reversal invariance at magnetic phase transitions, there is another type of systems where long-range spin ordering is suppressed by geometrical frustration. Usually, the geometrical frustration is caused by a combination of triangular-based exchange topology and antiferromagnetic interactions between neighbour spins. These systems fluctuate between degenerated ground states in the so called "spin-liquid" phase, and can stay in this regime down to very low temperatures. The absence of long-range dipole magnetic ordering does not necessarily mean that spin-liquids keep time-reversal invariance. Some other spin quantities such as toroidal moment ($\bf T$) and scalar spin chirality ($\chi $) have been proposed as possible candidates for the time-reversal breaking order parameters as well \cite{ref:4,ref:5,ref:5a,ref:6}. In spite of considerable efforts to find systems with actual realization of the spontaneous long-range toroidal or chiral order, in the absence of dipole ordering, only very recently a possible candidate of the latter has been found after observation of anomalous Hall effect in the metallic pyrochlore Pr$_2$Ir$_2$O$_7$ \cite{ref:7,ref:7a}.\\
\indent In insulating systems, however, identification of the long-range chiral ordering is more challenging but is not hopeless due to a possible coupling of the chiral order parameter to other measurable physical quantities. In this letter, we discuss this possibility for the geometrically frustrated antiferromagnets $R$BaMe$_4$O$_7$ ($R$=rare earth or Y, Me=Co,Fe) with a polar hexagonal lattice. Based on a simple symmetry consideration, we show that the polar nature of the crystal structure allows an appropriate coupling of the spin-chiral order parameter to spontaneous magnetization ($\bf M$) along the hexagonal axis. We propose this phenomenon as an explanation for the weak ferromagnetic component recently observed by Valldor et al. \cite{ref:8} in the spin-liquid YBaCo$_3$FeO$_7$.\\
\indent The prototype crystal structure of the YBaCo$_3$FeO$_7$ composition like in other members of the $R$BaMe$_4$O$_7$ family has hexagonal $P6_3mc$ symmetry \cite{footnote} and consists of corner-sharing MeO$_4$ tetrahedra arranged in alternating Kagome and triangular layers stacked along the $c$ axis (Fig. \ref{fig:F1}) \cite{ref:9,ref:10}. The Me-sublattice forms a new exchange topology where geometrical frustration is inherent to the presence of both trigonal bipyramids and triangular clusters \cite{ref:11,ref:12}.\\
\indent In spite of extremely strong exchange interactions in YBaCo$_3$FeO$_7$ (paramagnetic Curie-Weiss temperature is $\sim $ -2000K), no long-range magnetic ordering has been observed down to 1K \cite{ref:8}. It should be pointed out that some of the $R$BaCo$_4$O$_7$ compounds exhibit spin ordering below $\sim$ 100K, which has been shown to be closely related to a strong magnetoelastic coupling reducing the crystal structure symmetry down to monoclinic $P2_1$ \cite{ref:13}. Spin dynamics probed by high-resolution inelastic neutron scattering revealed a freezing below T$_f \sim$50K \cite{ref:8}, thus in a wide temperature range above T$_f$, YBaCo$_3$FeO$_7$ is in a spin-liquid state. Surprisingly, a very weak ferromagnetic component ($\sim 10^{-3}\mu _B$ per Co/Fe ion) along the $c$ axis and a hysteretic behaviour of the magnetization in magnetic field have been revealed below 590K. The existence of the small spontaneous magnetization in polycrystalline samples with the YBa$_{0.5}$Ca$_{0.5}$Co$_4$O$_7$ (below 387K \cite{ref:14}) and YBaCo$_4$O$_7$ (at room temperature \cite{ref:15,ref:16}) compositions had been reported before as well, indicating that this remarkable phenomenon is not a unique property of the Fe doped crystal. However, the higher transition temperature and the fact that the weak ferromagnetic component is observed in very clean single crystals and is anisotropic makes the YBaCo$_3$FeO$_7$ case a more compelling one. These observations along with numerous neutron diffraction measurements proving the lack of a coherent magnetic Bragg scattering above 100K \cite{ref:8,ref:11,ref:13,ref:17,ref:18} directly indicate that the time-reversal symmetry in the pure $R$BaCo$_4$O$_7$ or Fe doped compositions is broken without dipole spin ordering. The fundamental question immediately arises, what breaks the time-reversal invariance in all these materials?\\
\indent As it was mentioned above, a spin-dipole ordering is not the only order parameter which breaks time-reversal symmetry. Hypothetically, toroidal moment or scalar chirality are also considered as the possible variants of the time-odd order parameters. The latter is a three-spin quantity defined as a mixed product of spins $\chi = \bf {S_i} \cdot \left [ \bf {S_j} \times \bf {S_k} \right ]$ in a triangular plaquette. The scalar chirality takes a nonzero value for non-coplanar spin states and its sign represents whether the non-coplanar structure is either right- or left-handed. In spite of the time-odd nature of both $\bf T$ and $\chi $ order parameters, they have different transformational properties and therefore different types of coupling to macroscopic quantities. From symmetry point of view, the spin chirality defined above is a pseudoscalar i.e. it changes sign at a mirror reflection and stays invariant under rotations as it is demonstrated in Fig. \ref{fig:F2}, where a result of reflection by the mirror planes and the tree-fold rotation of the $P6_3mc$ space group are shown. In group theoretical language, it means that the chiral order parameter transforms as the time-odd $\Gamma ^-_2$ irreducible representation of the $P6_3mc1'$ space group (see Table \ref{table:irrep}, we use Miller and Love notation for the irreducible representations as it is implemented into ISOTROPY program \cite{ref:19}, the superscript indicates parity with respect to time inversion). A ferromagnetic component along the 6-fold axis ($M_z$) also transforms accordingly this representation and can  therefore be linearly coupled to $\chi $. Taking into account that magnetization is a vector and $\Gamma^-_2 \otimes \Gamma^-_2 \ni \Gamma^+_1$ and polarization along the $c$ axis ($P_z$) transforms as $\Gamma_1^+$, the general form of the free-energy invariant relating the macroscopic quantities can be written as $\chi \bf M \bf P$. Thus, scalar chirality can be coupled to magnetization only in polar crystals and in a general case, where polarization direction is not fixed by symmetry, a condensation of the chiral order parameter must induce magnetization in the direction of the polarization. In the case where polarization is frozen like in YBaCo$_3$FeO$_7$ (or other members of the family), coupling between $\chi $ and $\bf M$ is pseudoproper and the critical behaviour of both order parameters is identical. This offers the unique possibility to directly measure the critical exponent of the chiral order parameter with sufficient accuracy to verify the universality class of this kind of phase transitions.\\
\begin{table}[t]
\caption{Matrix of irreducible representations (irreps) for generators of $P6_3mc1'$ space group associated with wave vector {\bf k}=0 and physical quantiles transforming according to these representations. $t$ is the time-reversal operator.}
\centering 
\begin{tabular*}{0.48\textwidth}{@{\extracolsep{\fill}} c c c c c} 
\hline\hline\\ 
Irrep & $\lbrace C_{6+} | 0 0 \frac{1}{2} \rbrace$ & $\lbrace m_{xz} | 0 0 \frac{1}{2} \rbrace$ & $t$ & Quantity \\ [1.5ex] 
\hline\\ 
$\Gamma_1^+$ & 1 & 1 & 1 & $P_z$ \\ 
$\Gamma_1^-$ & 1 & 1 & -1 & $T_z$ \\
$\Gamma_2^-$ & 1 & -1 & -1 & $M_z, \chi $ \\
$\Gamma_5^-$ & $\left [ \begin{array}{cc} -\frac{1}{2} & \frac{\sqrt{3}}{2} \\ -\frac{\sqrt{3}}{2} & -\frac{1}{2} \end{array} \right ]$  & $\left [ \begin{array}{cc} -\frac{1}{2} & \frac{\sqrt{3}}{2} \\ \frac{\sqrt{3}}{2} & \frac{1}{2} \end{array} \right ]$ & $\left [ \begin{array}{cc} -1 & 0 \\ 0 & -1 \end{array} \right ]$ & $\eta_1,\eta_2$ \\
$\Gamma_6^-$ & $\left [ \begin{array}{cc} \frac{1}{2} & -\frac{\sqrt{3}}{2} \\ \frac{\sqrt{3}}{2} & \frac{1}{2} \end{array} \right ]$  & $\left [ \begin{array}{cc} \frac{1}{2} & -\frac{\sqrt{3}}{2} \\ -\frac{\sqrt{3}}{2} & -\frac{1}{2} \end{array} \right ]$ & $\left [ \begin{array}{cc} -1 & 0 \\ 0 & -1 \end{array} \right ]$ & $T_x,T_y$ \\
\\
\hline
\hline  
\end{tabular*}
\label{table:irrep} 
\end{table}
\indent In-plane ($T_x$,$T_y$) and out-of-plane ($T_z$) components of the macroscopic toroidal moment transform as the time-odd $\Gamma^-_6$ and $\Gamma^-_1$ representations, respectively (Table \ref{table:irrep}) and cannot form an appropriate coupling invariant linear in respect of $M_z$.\\
\indent The absence of both spatial and time inversions creates sufficient conditions for a linear magnetoelectric effect. The magnetic point group in the chiral phase of $R$BaCo$_4$O$_7$ is $6m'm'$ (magnetic space group is $P6_3m'c'$) which allows $\alpha _{11} = \alpha _{22}$ and $\alpha _{33}$ components of the magnetoelectric tensor to be non-zero. Thus, measurement of the magnetization in electric field or polarization in magnetic field can be an additional probe of a long-range chiral order in insulating crystals.\\
\indent The presented above symmetry-based phenomenological consideration is not able to reveal the microscopic origin of the phase transition to the chiral phase. However, let us discuss symmetry aspects of the possible interactions favouring non-coplanar spin arrangement in the case of well defined local correlations. First, we note that antisymmetic Dzyaloshinskii-Moriya (DM) interactions can in some circumstances create conditions for a non-zero scalar chirality in a spin triangle as it shown in Fig. \ref{fig:F3} (left). The phenomenological free-energy term responsible for these interactions is written in the form of $\bf {D_{ij}} \cdot \left [ \bf {S_i} \times \bf {S_j} \right ]$ where $\bf {D_{ij}}$ is the so called Dzyaloshinskii vector whose direction depends on the crystal symmetry \cite{ref:21}. In particular, in the hexagonal lattice of $R$BaCo$_4$O$_7$, it must be parallel to the mirror planes bisecting $ij$ bonds and can be decomposed into two components; out-of-plane $\bf {D_{out}}$ (parallel to the $c$ axis) and in-plane $\bf {D_{in}}$ (perpendicular to the $c$ axis). Due to the polar nature of the structure, the out-of-plane component is always either up or down in the whole crystal. This component removes the degeneracy between the two planar spin configurations with different vector chirality (defined as a vector product of the two neighboring spins, averaged over three spin pairs $\frac{2}{3\sqrt{3}} \sum_{\left \langle i,j \right \rangle} \left [ \bf {S_i} \times \bf {S_j} \right ]$, summation is taken over three pairs of sites along the sides of the triangle in an anti-clockwise direction) shown in Fig. \ref{fig:F3}a. Then, the  $\bf {D_{in}}$ component deviates the spins from the ($ab$) plane for the case when $\sum_{\left \langle i,j \right \rangle} \left [ \bf {S_i} \times \bf {S_j} \right ]\uparrow \uparrow \bf {D_{out}}$  and leaves them coplanar when $\sum_{\left \langle i,j \right \rangle} \left [ \bf {S_i} \times \bf {S_j} \right ]\uparrow \downarrow  \bf {D_{out}}$. Thus, depending on the direction of $\bf {D_{out}}$, the antisymmetric DM term can select non-coplanar spin configurations.\\ 
\indent The case, when $\bf {D_{in}}$ does not couple a $z$-component to the spins in a triangle (Fig. \ref{fig:F3}b (right)), is of particular interest. It is easy to see that admixture of the out-of-plane spin component does not change the $m'$ symmetry of the planar spin configuration and therefore can be added to it (Fig. \ref{fig:F3}c (right)). The coupling, however, cannot be linear since the $z$-component is invariant in respect of the three-fold axis, while the planar spin configuration is not. So the out-of-plane component has the higher symmetry ($3m'$) and can be coupled only as a linear-cubic term between the associated order parameters due to the necessary invariance in respect of the time inversion. Thus, interactions which relate the in-plane and out-of-plane spin components in this case are distinct from DM term coupling orthogonal components with identical transformational properties (bilinear coupling).\\
\begin{table}[t]
\caption{Symmetrized combinations of the atomic spins, $S$, on the Wyckoff position $6c$ (Kagome layer) transforming as components of the two-dimensional $\Gamma^-_5$ order parameter, $\eta_i (i=1,2)$ \cite{ref:19}.}
\centering 
\begin{tabular*}{0.48\textwidth}{@{\extracolsep{\fill}} c c c c c c c c} 
\hline\hline\\ 
$\eta_i$ & $S$ & {1\footnotemark[1]} & 2 & 3 & 4 & 5 & 6   \\ [1.5ex] 
\hline\\ 
& $S^x$ & 1 & -1 & 0 & 0 & -1 & 1 \\
$\eta_1$ & $S^y$ & 0 & -1 & 1 & -1 & 0 & 1 \\
& $S^z$ & 0 & 0 & 0 & 0 & 0 & 0 \\[1.5ex]
\hline\\ 
& $S^x$ & $-\frac{1}{\sqrt{3}}$ & $-\frac{1}{\sqrt{3}}$ & $\frac{2}{\sqrt{3}}$ & $-\frac{2}{\sqrt{3}}$ & $\frac{1}{\sqrt{3}}$ & $\frac{1}{\sqrt{3}}$ \\[1.5ex]
$\eta_2$ & $S^y$ & $-\frac{2}{\sqrt{3}}$ & $\frac{1}{\sqrt{3}}$ & $\frac{1}{\sqrt{3}}$ & $-\frac{1}{\sqrt{3}}$ & $\frac{2}{\sqrt{3}}$ & $-\frac{1}{\sqrt{3}}$ \\[1.5ex]
& $S^z$ & 0 & 0 & 0 & 0 & 0 & 0 \\[1.5ex]
\hline\hline 
\end{tabular*}
\footnotetext[1]{1-($x,-x,z$), 2-($x,2x,z$), 3-($-2x,-x,z$), 4-($2x,x,z+\frac{1}{2}$), 5-($-x,x,z+\frac{1}{2}$), 6-($-x,-2x,z+\frac{1}{2}$); $x\sim \frac{1}{6}, z\sim \frac{11}{16}$}
\label{table:fun} 
\end{table}
\indent More generally, the global distortions of the type shown in Fig. \ref{fig:F3}a (right) keeping translational invariance of the crystal can be presented as combinations of basis functions of the $\Gamma_5^-$ irreducible representation of the $P6_3mc1'$ space group (Table \ref{table:fun}). The associated two-dimensional order parameter $(\eta_1,\eta_2)$ forms a general coupling invariant with the out-of-plane ferromagnetic component, $3M_z \eta_1^2 \eta_2-M_z\eta_2^3$, which is reduced down to the simple linear-cubic term, $M_z\eta_2^3$, for the $(0,\eta_2)$ direction. The latter is the equilibrium order parameter for the most symmetric variant of the spin structure with the vector chirality shown in Fig. \ref{fig:F3}a (right). Minimization of the free-energy part, $F=\gamma M_z^2+ \beta M_z\eta^3_2$, in respect of $M_z$, gives the non-zero equilibrium  value for the latter, $M_z=-\frac{\beta \eta^3_2}{2\gamma}$. Thus, in spite of the passive role of the DM interactions, the macroscopic magnetization and scalar chirality are expected to be non-zero due to the existence of the linear-cubic type of the coupling. The chiral free-energy term, $\chi \bf {MP}$, (which shows that the system gains energy in the case of the non-coplanar spin arrangement) can clearly provide this coupling with some "adopted" to local symmetry meaning of the $\bf M$ and $\bf P$ quantities and in a general case it represents antisymmetric interactions distinct from DM. These interactions imply a "chiral dependent" part of the exchange Hamiltonian, tending to bring three spins to the mutually orthogonal orientation. The local meaning of $\bf M$ and $\bf P$ adopted for a single triangular unit in a structure should ensure vanishing of the coupling invariant for some spin configurations whose symmetry forbids non-coplanar states as in the case of $3m$ point group shown in Fig. \ref{fig:F3}c (left). In some particular cases, the DM exchange can contribute to the free-energy chiral term too (as it was shown above), since the existence of the bilinear coupling invariant between order parameters means existence of a linear-cubic one as well. To further clarify the relevant interactions responsible for the chiral state in $R$BaMe$_4$O$_7$, numerical simulations based on the Monte-Carlo algorithm with different terms in magnetic Hamiltonian including the chiral-dependent $\chi \bf {MP}$ one are desirable.\\
\indent In conclusion, based on symmetry consideration, we argue that the weak ferromagnetic component, measured in YBaCo$_3$FeO$_7$ and some other isostructural compositions, in spin-liquid regime, is related to a long-range chiral ordering. The coupling between both macroscopic physical quantities is possible in polar crystals due to a phenomenological trilinear invariant of the form $\chi \bf {MP}$. This invariant, in a general case represents interactions distinct from Dzyaloshinskii-Moriya ones and provides a linear-cubic coupling between orthogonal spin-components. Due to the frozen nature of the polarization in the hexagonal lattice of $R$BaCo$_4$O$_7$, the coupling between $\chi $ and $\bf M$ is pseudoproper and the critical behaviour of both order parameters is identical. This offers a unique possibility to identify the universality class of these exotic phase transitions. The obtained result here provides a physical ground for a search of other possible systems where spin chirality is the primary order parameter.

\end{document}